\documentclass[aps,showkeys,amsmath,amssymb]{revtex4}
\usepackage{bm}

\begin{document}

\title{Some Further Properties of the Accelerated Kerr-Schild Metrics}
\author{Metin G{\" u}rses}
\email{gurses@fen.bilkent.edu.tr}
\affiliation{Department of Mathematics, Faculty of Sciences,\\
             Bilkent University, 06533 Ankara, Turkey}

\author{{\" O}zg{\" u}r Sar{\i}o\u{g}lu}
\email{sarioglu@metu.edu.tr}
\affiliation{Department of Physics, Faculty of Arts and Sciences,\\
             Middle East Technical University, 06531 Ankara, Turkey}

\date{\today}

\begin{abstract}
We extend the previously found accelerated Kerr-Schild metrics for
Einstein-Maxwell-null dust and Einstein-Born-Infeld-null dust
equations to the cases including the cosmological constant. This
way we obtain the generalization of the charged de Sitter metrics in static
space-times. We also give a generalization of the zero acceleration limit
of our previous Einstein-Maxwell and Einstein-Born-Infeld solutions.
\end{abstract}


\maketitle

\section{\label{intro} Introduction}

Using a curve $C$ in $D$-dimensional Minkowski space-time $M_{D}$,
we have recently studied the Einstein-Maxwell-null dust \cite{gur1} and
Einstein-Born-Infeld-null dust field equations \cite{gur2},
Yang-Mills equations \cite{ozg}, and Li{\' e}nard-Wiechert potentials
in even dimensions \cite{gur3}. In the first three works we found 
some new solutions generalizing the Tangherlini \cite{tan}, 
Pleba{\' n}ski \cite{pleb}, and Trautman \cite{trt} solutions, 
respectively. The last one proves that the accelerated scalar or 
vector charged particles in even dimensions lose
energy. All of the solutions contain a function $c$ which is
assumed to depend on the retarded time $\tau_{0}$ and all
accelerations $a_{k}, (k=0,1,2, \cdots)$, see
\cite{gur1}-\cite{gur3}. We also assumed that when the motion is
uniform or the curve $C$ is a straight line in $M_{D}$, this
function reduces to a function depending only on the retarded time
$\tau_{0}$. In this work we first relax this assumption and give
the most general form of the function $c$ when the curve $C$ is
a straight line. In addition, we also generalize our previous
accelerated Kerr-Schild metrics by
including the cosmological constant in arbitrary $D$-dimensions
for the Einstein-Maxwell and in four dimensions for the 
Einstein-Born-Infeld theories. The solutions presented here 
can be interpreted as the solutions of the Einstein-Maxwell-null 
perfect fluid field equations with a constant pressure $\Lambda$ 
or Einstein-Maxwell-null dust field
equations with a cosmological constant $\Lambda$.
In our treatment we adopt the second interpretation.

Our conventions are similar to the conventions of our earlier works  
\cite{gur1}, \cite{gur2}, \cite{gur3}. In a $D$-dimensional Minkowski
space-time $M_{D}$, we use a parameterized curve $C= \{x^{\mu}
\in M_{D}: x^{\mu}=z^{\mu}(\tau)\, , \mu=0,1,2, \cdots , D-1 \, , 
\tau \in I \}$
such that $\tau$ is a parameter of the curve and $I$ is an
interval on the real line ${\mathbb R}$. We define the world
function $\Omega$ as
\begin{equation}
\Omega=\eta_{\mu \nu}\,(x^{\mu}-z^{\mu}(\tau))\,(x^{\nu}-z^{\nu}(\tau)),
\label{dist}
\end{equation}
where $x^{\mu}$ is a point not on the curve $C$.
There exists a point $z^{\mu}(\tau_{0})$ on the non-spacelike curve $C$ 
which is also on the light cone with the vertex located at the point
$x^{\mu}$, so that $\Omega(\tau_{0})=0$. Here $\tau_{0}$ is the
retarded time. By using this property we find that
\begin{equation}
\lambda_{\mu} \equiv \partial_{\mu}\, \tau_{0} =
\frac{x^{\mu}-z^{\mu}(\tau_{0})}{R}
\end{equation}
where 
$R \equiv \dot{z}^{\mu}(\tau_{0})\,(x_{\mu}-z_{\mu}(\tau_{0}))$ is the
retarded distance. Here a dot over a letter denotes
differentiation with respect to $\tau_{0}$. It is easy to show
that $\lambda_{\mu}$ is null and satisfies
\begin{equation}
\lambda_{\mu, \nu}=\frac{1}{R}\, [\eta_{\mu \nu}-\dot{z}_{\mu}\,
\lambda_{\nu}-\dot{z}_{\nu}\, \lambda_{\mu}-(A-\epsilon)\,
\lambda_{\mu}\, \lambda_{\nu}]
\end{equation}
where $A \equiv \ddot{z}^{\mu}\,
(x_{\mu}-z_{\mu}(\tau_{0}))$ and $\dot{z}^{\mu}\,
\dot{z}_{\mu}=\epsilon= -1,0$. Here $\epsilon=-1$ and $\epsilon=0$
correspond to the time-like and null velocity vectors,
respectively. One can also show explicitly that $\lambda^{\mu}\,
\dot{z}_{\mu}=1$ and $\lambda^{\mu}\, R_{,\, \mu}=1$. Define $a
\equiv \frac{A}{R}=\lambda^{\mu} \ddot{z}_{\mu}$, then
\begin{equation}
\lambda^{\mu}\, a_{, \, \mu}=0.
\end{equation}
Furthermore defining (letting $a_{0}=a$)
\begin{equation}
a_{k} \equiv \lambda_{\mu}\, 
\frac{d^{k+2} \, z^{\mu}(\tau_{0})}{d \tau_{0} ^{k+2}}, \;\; 
k=0,1,2, \cdots \label{aks}
\end{equation}
one can show that
\begin{equation}
\lambda^{\mu}\, a_{k,\, \mu}=0,\;\;\; \forall k=0,1,2, \cdots \, .
\end{equation}
Hence any function $c$ satisfying
\begin{equation}
\lambda^{\mu}\, c_{,\, \mu}=0, \label{con1}
\end{equation}
has arbitrary dependence on all $a_{k}$'s and
$\tau_{0}$. Using the above curve kinematics, we showed that
Einstein-Maxwell-perfect fluid equations with the Kerr-Schild
metric give us the following result

{\bf Proposition 1}. Let the space-time metric and the
electromagnetic vector potential be respectively given by 
$g_{\mu \nu}=\eta_{\mu\nu}-2V \lambda_{\mu} \lambda_{\nu},\,\, 
A_{\mu}=H\, \lambda_{\mu}$, where $V$ and $H$ are some differentiable
functions in $M_{D}$. Let $V$ and $H$ depend on $R$, $\tau_{0}$
and functions $c_{i} \, (i=1,2, \cdots)$ that satisfy
(\ref{con1}), then the Einstein equations reduce to the following
set of equations [see \cite{gur1} for details] 
\begin{eqnarray}
\kappa p+\Lambda & = & \frac{1}{2} V^{\prime \prime}+
\frac{3D-8}{2R} V^{\prime}
+\frac{(D-3)^2}{R^2} V, \label{pres1}\\
\kappa (H^{\prime})^2 & = & V^{\prime \prime}+\frac{D-4}{R} V^{\prime}-
\frac{2V}{R^2}(D-3), \label{denk1}\\
\kappa (p+\rho) & = & q-\kappa \eta^{\alpha \beta}\, H_{,\alpha} H_{,\beta}
\nonumber\\
& & + \, 2\, \left[ \frac{2(A-\epsilon)(D-3)V}{R^2}-
\sum_{i=1}(w_{i} \, c_{i, \alpha}\, \dot{z}^{\alpha}) \right] , \label{rho1}\\
\sum_{i=1}\, w_{i} \, c_{i, \alpha} & = & \left[\sum_{i=1}\, (w_{i}\,
c_{i, \beta} \,\dot{z}^{\beta}) \right] \, \lambda_{\alpha},
\label{cler1}
\end{eqnarray}
where
\begin{equation}
w_{i}=V^{\prime}_{,c_{i}}+\frac{D-4}{R}\, V_{,c_{i}}-
\kappa H^{\prime}\, H_{,c_{i}} ,
\end{equation}
and prime over a letter denotes partial differentiation
with respect to $R$. Here $\kappa$ is the gravitational constant,
$p$ and $\rho$ are, respectively, the pressure and energy density
of the fluid, $\Lambda$ is the cosmological constant and the function
$q$ is defined by
\[
q=\eta^{\alpha \beta}\, V_{, \alpha \beta}-\frac{4}{R}\dot{z}^{\alpha}\, 
V_{,\alpha}
+2(\epsilon-A)\, \frac{\lambda^{\alpha} V_{, \alpha}}{R}+
[2\epsilon (-D+3)+2A(D-2)]\, \frac{V}{R^2}.
\]
Please refer to \cite{gur1} for this Proposition.

For the case of the Einstein-Born-Infeld field equations with
similar assumptions, we have the following proposition (please see
\cite{gur2} for the details of the Proposition)

{\bf Proposition 2}. Let $V$ and $H$ depend on $R$,
$\tau_{0}$ and functions $c_{i} \, (i=1,2, \cdots)$ that satisfy
(\ref{con1}), then the Einstein equations reduce to the following
set of equations
\begin{eqnarray}
\kappa p+\Lambda & = & V^{\prime \prime}+\frac{2}{R} V^{\prime}-
\kappa b^2\, [1-\Gamma_{0}], \label{pres2}\\
\kappa \frac{(H^{\prime})^2}{\Gamma_{0}}
& = & V^{\prime \prime}-\frac{2V}{R^2} , \label{denk2}\\
\kappa (p+\rho) & = & \sum_{i=1}\, \left[ V_{,c_{i}}\, 
(c_{i, \alpha}\,^{,\alpha})-\frac{4}{R}
V_{,c_{i}}\,(c_{i,\alpha}\dot{z}^{\alpha})  \right. \nonumber \\
& & \left. -\frac{\kappa}{\Gamma_{0}}\, (H_{,c_{i}})^2 \,(c_{i,\alpha}\,
c_{i}\,^{,\alpha}) \right ]-\frac{2A}{R}
\left( V^{\prime}-\frac{2V}{R} \right) , \label{pres3}\\
\sum_{i=1}\, w_{i}\,c_{i, \alpha} & = & \left[\sum_{i=1}\, (w_{i}\, c_{i,
\beta} \,\dot{z}^{\beta}) \right]\, \lambda_{\alpha}, \label{cler2}
\end{eqnarray}
where
\begin{equation}
w_{i}=V^{\prime}_{,c_{i}}-
\frac{\kappa H^{\prime}}{\Gamma_{0}}\,H_{,c_{i}} , \;\;\;
\Gamma_{0}=\sqrt{1-\frac{(H^{\prime})^2}{b^2}},
\end{equation}
and prime over a letter denotes partial
differentiation with respect to $R$. Here $\kappa (=8\pi)$ is 
the gravitational constant,
$p$ and $\rho$ are, respectively the pressure and the energy density 
of the fluid,
$b$ is the Born-Infeld parameter (as $b \rightarrow \infty$ one 
arrives at the Maxwell
limit), and $\Lambda$ is the cosmological constant.

\section{\label{sect2} Null-Dust Einstein-Maxwell Solutions in $D$-Dimensions
with a Cosmological Constant}

In this section we assume zero pressure. Due to
the existence of the cosmological constant $\Lambda$, one may
consider this as if there is a constant pressure as the source of 
the field equations. We shall not adopt this interpretation. 
Instead, we think of this as if there is a null dust,
a Maxwell field and a cosmological constant as the source of 
the Einstein field equations. Hence assuming that the null fluid 
has no pressure in Proposition 1, we have the following result:

{\bf Proposition 3}. Let $p=0$. Then
\begin{eqnarray}
V&=& \left \{\begin{array}{ll} \frac{\kappa e^2\,(D-3)}{2(D-2)}
R^{-2D+6}+m R^{-D+3}+\frac{\Lambda}{(D-2)(D-1)}\, R^2,
 & (\mbox{$ D \ge 4$})\\
-\frac{\kappa}{2} e^2 \, \ln{R+m} +\frac{\Lambda}{2}\,R^2, 
&(\mbox{$D=3$})
\end{array}
\right. \label{e50}\\
H&=& \left \{\begin{array}{ll}
c+\epsilon \,e\, R^{-D+3},  & (\mbox{$D \ge 4$})\\
c+\epsilon e\, \ln{R} .  & (\mbox{$D=3$})
\end{array}
\right. \label{e51}
\end{eqnarray}
The explicit expressions of the energy density $\rho$ and the current 
vector $J_{\mu}$
do not contain the cosmological constant $\Lambda$ and are identical with the
ones given in \mbox{\cite{gur1}}, so we don't rewrite those long formulas here.

Here $M=m+\epsilon\,\kappa (3-D) e c$ for $D \ge 4$ and
$M=m+\frac{\kappa}{2}\,e^2 +\epsilon\, \kappa e c$ for $D=3$. In
all cases $e$ is assumed to be a function of $\tau_{0}$ only but
the functions $m$ and $c$ which are related through the arbitrary
function $M(\tau_{0})$ (depends on $\tau_{0}$ only) do depend on
the scalars $a_{k} \, (k \ge 0)$. Of course $\Lambda$ is any real number.

In Proposition 3  we have chosen the integration
constants ($R$ independent functions) as the functions $c_{i}$
($i=1,2,3$) so that $c_{1}=m$, $c_{2}=e$, $c_{3}=c$ and
\begin{equation}
c=c(\tau_{0}, a_{k}), \;\; e=e(\tau_{0}),
\end{equation}
\begin{eqnarray}
m=\left \{ \begin{array}{ll}
M(\tau_{0})+\epsilon\,\kappa (D-3) e c , & (\mbox{$D \ge 4$}) \label{md41}\\
M(\tau_{0})-\frac{\kappa}{2}\,e^2-\epsilon\, \kappa e c , & (\mbox{$D=3$})
\label{md31}
\end{array}
\right.
\end{eqnarray}
where $a_{k}$'s are defined in (\ref{aks}).

{\bf Remark 1}. We can have pure null dust solutions when $e=0$. 
The function $c$ in this case can be gauged away, that is we can 
take $c=0$. The Ricci tensor takes its simplest form 
$R_{\mu \nu}=\rho \lambda_{\mu}\, \lambda_{\nu}+
\Lambda\, g_{\mu \nu}$ then. In this case we have
\begin{eqnarray}
V & = & m R^{3-D}+\frac{\Lambda}{(D-2)(D-1)}\,R^2 , \nonumber \\
\rho & = & \frac{2-D}{\kappa}\, [a (1-D)m+\dot{m}]\, R^{2-D}
\end{eqnarray}
for $D \ge 4$ and
\begin{eqnarray}
V & = & m+ \frac{\Lambda}{2} \, R^2, \nonumber \\
\rho & = & \frac{2ma -\dot{m}}{\kappa R}
\end{eqnarray}
for $D=3$. Such solutions are usually called as the {\it Photon Rocket}
solutions \cite{kin}, \cite{bon}. We give here the $D$ dimensional
generalizations of this type of metrics with a cosmological constant.

{\bf Remark 2}. If  $e=m=0$ we obtain a metric
\[
g_{\mu \nu} = \eta_{\mu \nu}- \frac{2\Lambda \, R^2}{(D-1)(D-2)}\, \,
 \lambda_{\mu}\, \lambda_{\nu}, \;\;\, (D \ge 3)
 \]
which clearly corresponds to the $D$-dimensional de Sitter space.

{\bf Remark 3}. The static limit $a_{0} \equiv a=0$ of our solutions
with a constant $c$ are the charged Tangherlini solutions with
a cosmological constant. If the function $c$ is not chosen to be
a constant, we obtain their  generalizations (see Section \ref{sect4}).

\section{\label{sect3} Null-Dust Einstein-Born-Infeld Solutions in 
4-Dimensions with a Cosmological Constant}

Using Proposition 2, and assuming zero pressure, we find the complete
solution of the field equations.

{\bf Proposition 4}. Let $p=0$. Then
\begin{eqnarray}
V & = & \frac{m}{R}-4 \pi e^2 \frac{F(R)}{R}+\frac{\Lambda}{6}\, R^2 , \\
H & = & c-\epsilon\,e \int^{R}\, \frac{dR}{\sqrt{R^4+e^2/b^2}} ,
\end{eqnarray}
where
\begin{eqnarray}
m & = & M(\tau_{0})+ 8\pi \epsilon e c, \label{mbi41}\\
F(R) & = & \int^{R}\, \frac{dR}{R^2+\sqrt{R^4+e^2/b^2}} .
\end{eqnarray}
Here $e$ is assumed to be a function of $\tau_{0}$ only
but the functions $m$ and $c$ which are related through the
arbitrary function $M(\tau_{0})$ do depend on the scalars 
$a_{k}$, ($k \ge 0$). 

We have chosen the integration constants ($R$ independent functions) 
as the functions $c_{i}$ ($i=1,2,3$) so that $c_{1}=m$, $c_{2}=e$, 
$c_{3}=c$ and
\[
c=c(\tau_{0},a_{k}), \;\; e=e(\tau_{0}), \;\; 
m=M(\tau_{0})+8 \pi \epsilon e c .
\]

{\bf Remark 4}. In the static limit with a constant $c$, we obtain the 
Pleba{\' n}ski solution with a cosmological constant \cite{fern}. If 
the function $c$ is not a constant, we can also give a class of 
solutions of the Einstein-Born-Infeld-null dust equations
with a cosmological constant (see Section \ref{sect4}).

\section{\label{sect4} Straight Line Limits}

When the accelerations $a_{k} \, (k \ge 0)$ vanish, the curve $C$ is a 
straight line in $M_{D}$.  In this limit we have the following: 
$\tau_{0}=t-r$, $z^{\mu}=n^{\mu}\, \tau_{0}$, $n^{\mu} \equiv (1,0,0,0)$, 
$\dot{z}^{\mu}=n^{\mu}$ and $R=-r$. Moreover,
\begin{equation}
x^{\mu}=(t, \vec{x}), \;\; \lambda_{\mu}=(1, -\frac{\vec{x}}{r}), \;\;
r^2=\vec{x} \cdot \vec{x}.
\end{equation}
In this case the function $c$ arising in the metrics introduced in the
previous sections can be assumed to depend on some other functions 
$\xi_{(I)}$ so that
$\lambda^{\mu}\, \xi_{(I), \mu}=0$ ($I=1,2, \cdots , D-2$)\, \cite{mar}. 
As an example let
$\xi_{(I)}=\vec{l}_{(I)} \cdot \frac{\vec{x}}{r}$, where $\vec{l}_{(I)}$ 
are constant vectors. It is easy to show that 
$\lambda^{\mu}\, \xi_{(I),\, \mu}=0$. 
Hence in this (straight line) limit we assume that $c=c(\tau_{0}, \xi_{(I)})$. 
From this simple example we may define more general functions satisfying our 
constraint equation $c_{,\mu}\, \lambda^{\mu}=0$. Let $X_{\mu}$ be 
a vector satisfying
\begin{equation}
X_{\mu, \,\nu}=b_{0}\, \eta_{\mu \nu}+b_{1}(k_{\mu}\, \lambda_{\nu}+
k_{\nu}\, \lambda_{\mu})+ b_{2}\, \lambda_{\mu}\, \lambda_{\nu}
+ Q_{\mu \nu},   \label{xi}
\end{equation}
where $b_{0}, b_{1}, b_{2}$ are some arbitrary functions, $k_{\mu}$ is 
any vector and $Q_{\mu \nu}$ is any antisymmetric tensor in $M_{D}$. Then 
any vector $X_{\mu}$ satisfying (\ref{xi}) defines a scalar 
$\xi=\lambda^{\mu}\, X_{\mu}$ so that $\lambda^{\mu}\, \xi_{,\, \mu}=0$. 

The simple example given at the beginning of this section corresponds to 
a constant vector, $X_{\mu}=l_{\mu}=(l_{0}, \vec{l})$. Hence, $\xi$
becomes a function of the spherical angles. For instance, in four dimensions,
$\xi= l_{0}+l_{1}\, \cos{\phi} \, \sin{\theta}+l_{2}\, 
\sin{\phi} \, \sin{\theta}+ l_{3}\, \cos{\theta}$  where 
$l_{0}, l_{1}, l_{2}$, and $l_{3}$ are the constant
components of the vector $l_{\mu}$. In the straight
line limit, the metric can be transformed easily to the form
\begin{equation}
ds^2 = -(1+2V)\, dT^2+\frac{1}{1+2V}\, dr^2+r^2\, d \Omega_{D-2}^2,
\end{equation}
where
\[
dT=dt- \frac{2V\, dr}{1+2V},
\]
and $d\Omega_{D-2}^2$ is the metric on the $D-2$-dimensional unit sphere.
The above form of the metric is valid both for the Einstein-Maxwell and for the
Einstein-Born-Infeld theories. For the case of the Einstein-Maxwell-null dust
with a cosmological constant we have
\begin{equation}
V= \left\{ \begin{array}{ll}\frac{\kappa e^2\,(D-3)}{2(D-2)}\,
r^{-2D+6}+m (-1)^{D+1}\,r^{-D+3}+
\frac{\Lambda}{(D-2)(D-1)}\, r^2, & \mbox{($ D \ge 4$)}\\
m-\frac{\kappa}{2} e^2 \, \ln{r}+\frac{\Lambda}{2}\,r^2,& \mbox{($D=3$)}
\end{array} \right.
\end{equation}
and the function $H$ defining the electromagnetic vector potential
is given by
\begin{equation}
H=\left\{ \begin{array}{ll} c+\epsilon \,e\,(-1)^{D+1}\, r^{-D+3},&
( \mbox{$D \ge 4$})\\
c+\epsilon e \, \ln{r} . &(\mbox{$D=3$})
\end{array} \right.
\end{equation}
This solution is a generalization of the Tangherlini solution \cite{tan}.
The relationship between $c$ and $m$ are given in (\ref{md41}), 
but in this case the function $c$ is a function of the scalars $\xi_{(I)}$ 
as discussed in the first part of this section. For the case of the 
Einstein-Born-Infeld-null dust with a cosmological constant, we have
\begin{eqnarray}
V & = & -\frac{m}{r}+4\pi e^2 \frac{F(r)}{r}+\frac{\Lambda}{6}\, r^2 , \\
H & = & c+\epsilon\,e \int^{r}\, \frac{dr}{\sqrt{r^4+e^2/b^2}}
\end{eqnarray}
where
\begin{eqnarray}
m & = & M(\tau_{0})+8\pi \epsilon e c, \label{mbi42}\\
F(r) & = & -\int^{r}\, \frac{dr}{r^2+\sqrt{r^4+e^2/b^2}}.
\end{eqnarray}
This solution is a generalization of the Pleba{\' n}ski et al \cite{pleb}
static solution of the Einstein-Born-Infeld theory. Our
generalization is with the function $c$ depending arbitrarily on the
scalars $\xi_{(I)}$.

{\bf Remark 5}. When the function $c$ is not a constant, 
the mass $m$ defined through the relations (\ref{md41}) 
or (\ref{mbi42}) is not a constant anymore, it depends on the
angular coordinates.

\section{\label{concl} Conclusion}

We have reexamined the accelerated Kerr-Schild geometries for two purposes.
One of them is to generalize our earlier solutions of Einstein-Maxwell-null 
dust \cite{gur1} and Einstein-Born-Infeld-null dust field equations \cite{gur2}
by including a cosmological constant. The other one is to generalize the static
limit (straight line limit) of the above mentioned solutions. Previously in the
static limit, we were assuming the function $c$ to be a constant. As long 
as this function satisfies the condition $\lambda^{\mu}\, c_{,\mu}=0$, 
as we have seen in Section \ref{sect4} (although the acceleration 
scalars $a_{k} \,(k \ge 0)$ are all zero) we can still obtain the 
generalization of the charged Tangherlini \cite{tan} and Pleba{\' n}ski 
\cite{pleb} solutions.

\begin{acknowledgments}

We thank Marc Mars for useful suggestions. This work is partially supported 
by the Scientific and Technical Research Council of Turkey and by the 
Turkish Academy of Sciences.

\end{acknowledgments}



\begin{thebibliography}{99}
\bibitem{mar} We would like to thank Marc Mars for drawing our
attention to such a generalization.
\end{thebibliography}
\end{document}